\documentclass[conference]{IEEEtran}
\IEEEoverridecommandlockouts
\usepackage{cite}
\usepackage{amsmath,amssymb,amsfonts}
\usepackage{algorithmic}
\usepackage{graphicx}
\usepackage{textcomp}
\usepackage{xcolor}
\usepackage{paralist}
\usepackage{listings}
\usepackage{verbatim}
\usepackage{tablefootnote}
\usepackage{float}
\usepackage{pdflscape}
\usepackage{hyperref}
\usepackage{cleveref}
\usepackage{subfigure}
\usepackage{xcolor}
\usepackage{float}
\usepackage[most]{tcolorbox}
\usepackage{multirow}
\usepackage{graphicx}

\usepackage{url}
\usepackage{booktabs}
\AtBeginDocument{%
  \providecommand\BibTeX{{%
    Bib\TeX}}}

\lstdefinelanguage{HTML}{
  morekeywords={html,head,title,body,div,span,h1,h2,h3,p, pre, code, img,script,style},
  sensitive=false,
  morecomment=[s]{<!--}{-->},
  morestring=[b]",
}

\newcommand{\new}[1]{{#1}}

\def\BibTeX{{\rm B\kern-.05em{\sc i\kern-.025em b}\kern-.08em
    T\kern-.1667em\lower.7ex\hbox{E}\kern-.125emX}}
\begin{document}

\title{What Challenges Do Developers Face When Using Verification-Aware Programming Languages?\\
}

\author{\IEEEauthorblockN{1\textsuperscript{st} Francisco Oliveira}
\IEEEauthorblockA{\textit{Faculty of Engineering} \\
\textit{University of Porto}\\
Porto, Portugal \\
}
\and
\IEEEauthorblockN{2\textsuperscript{nd} Alexandra Mendes}
\IEEEauthorblockA{\textit{INESC TEC, Faculty of Engineering, } \\
\textit{University of Porto}\\
Porto, Portugal \\
}
\and
\IEEEauthorblockN{3\textsuperscript{rd} Carolina Carreira}
\IEEEauthorblockA{\textit{Carnegie Mellon University} \\
\textit{INESC-ID \& IST, University of Lisbon}\\
Lisbon, Portugal \\
}

}
\maketitle

\begin{abstract}
Software reliability is critical in ensuring that the digital systems we depend on function correctly. In software development, increasing software reliability often involves testing. However, for complex and critical systems, developers can use Design by Contract (DbC) methods to define precise specifications that software components must satisfy. Verification-Aware (VA) programming languages support DbC and formal verification at compile-time or run-time, offering stronger correctness guarantees than traditional testing.
However, despite the strong guarantees provided by VA languages, their adoption remains limited. In this study, we investigate the barriers to adopting VA languages by analyzing developer discussions on public forums using topic modeling techniques. We complement this analysis with a developer survey to better understand the practical challenges associated with VA languages. Our findings reveal key obstacles to adoption, including steep learning curves and usability issues. Based on these insights, we identify actionable recommendations to improve the usability and accessibility of VA languages. Our findings suggest that simplifying tool interfaces, providing better educational materials, and improving integration with everyday development environments could improve the usability and adoption of these languages. Our work provides actionable insights for improving the usability of VA languages and making verification tools more accessible.
\end{abstract}


\section{Introduction}
In today's world, software is integral to almost every aspect of our lives, from personal communication and entertainment to critical systems like air traffic control and IoT devices. As software becomes increasingly prevalent, ensuring its reliability is essential.
The importance of reliability is even greater when considering critical applications where failures can be life-threatening and cause extensive damage~\cite{parnas_safety}, such as financial damage and risk users' privacy. %

Verification-aware (VA) languages and systems, such as Dafny~\cite{leino2010dafny}, VCC~\cite{cohen2009vcc}, %
Why3~\cite{filliatre2013why3}, and Verus \cite{lattuada2024verus}, go beyond traditional programming languages by integrating formal verification capabilities directly into the development process. These languages enable the verification of program correctness against user-defined specifications or contracts, using formal methods to mathematically prove that a program adheres to its specification. This built-in verification makes it possible to detect bugs early in the development cycle, offering a higher degree of assurance about software correctness and safety. %

Despite their importance, the adoption of VA languages is still low. The reason usually attributed to this is the level of expertise required from the developers and challenges that they often face when trying to prove a program correct. %
Previous work identifies some challenges when using these tools, such as the lack of counter-examples~\cite{Todorov201884}, lack of clarity of the error messages~\cite{schoolderman_is_2019}, and a general difficulty in understanding why proofs fail~\cite{gudemann_online_2021}. However, to the best of our knowledge, no previous studies have been made to systematically collect the challenges and struggles that VA language developers face.%

In this paper, we aim to identify 
challenges usually encountered when using VA languages 
with the aim of supporting VA language, verifier, and IDE designers to improve the developer experience and, hopefully contribute to a wider adoption. 

We conducted a literature review to identify challenges in the use of VA languages that are documented in the literature and used topic modeling, in particular Latent Dirichlet Allocation (LDA)~\cite{blei_latent_2003}, to extract challenges from 
developer posts on Stack Overflow.
This was followed by a user study to validate the identified challenges and improvements that can be made to VA languages to improve adoption and user experience. 

We address the following research questions:
\paragraph*{\textbf{RQ1}} What challenges do verification-aware language practitioners face?
\paragraph*{\textbf{RQ2}} What is the context of the first contact with verification-aware languages?
\paragraph*{\textbf{RQ3}} What improvements can be made to existing tools to overcome those challenges?

\vskip 0.1em
\noindent\noindent
{\bf Contributions.} The main contributions of this work are:
\begin{itemize}
    \item A collection of challenges that affect VA language developers, obtained through a literature review and the analysis of 1,420 questions from Stack Overflow, validated and complemented by a user study with 31 practitioners. %
    \item An overview of practitioners' first contact with VA languages.
    \item A collection of suggested improvements, validated by beginner and experienced VA language practitioners, that can inform language, verifier, and IDE designers. %
\end{itemize}

\begin{figure*}
    \centering
    \includegraphics[width=0.85\linewidth]{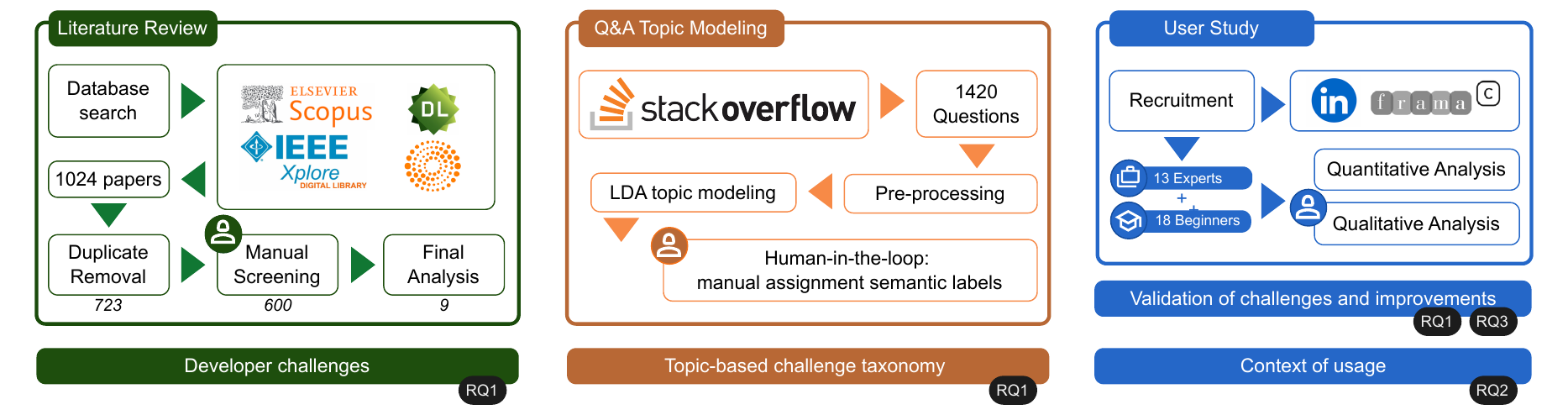}
    \caption{Overview of the three-step research methodology.
    }
    \label{fig:overview}
\end{figure*}

\section{Study Design}

We used a three-step approach as shown in \Cref{fig:overview}. In \emph{Step~1}, we performed a literature review to understand the context in which practitioners apply VA languages and the challenges that arise from their use. \emph{Step~2} consisted of topic modeling, in particular using Latent Dirichlet Allocation (LDA)~\cite{blei_latent_2003}, to extract challenges from 
developer
posts on Stack Overflow.
Finally, \emph{Step~3} consisted of a user-study to validate the challenges collected in the previous two steps, to explore first contexts of users' contact with these languages, and to explore potential improvements.
Steps 1, 2, and 3 contribute to answering RQ1. Step 3 contributes to answering questions RQ2 and RQ3. %

\subsection{Verification-Aware Languages Considered}\label{subsec:VAL_considered}
We consider as VA programming languages those that natively support constructs for automated program verification, such as pre-conditions, post-conditions, and invariants. For our search queries, we defined a set of VA languages to be considered, 
which resulted from a search of Verifiable Programming Languages used in industry taken from the Formal Methods Europe Industry Committee's website,\footnote{\url{https://fme-industry.github.io/}} consulted on 20/11/2023. 
The languages that did not comply with our definition of VA languages were removed.

Following a three-person analysis of the technologies listed, we arrived at the following set: Bandera~\cite{corbett2000bandera}, Dafny~\cite{leino2010dafny}, Escher C Verifier~\cite{escher2017ecv}, Eiffel~\cite{meyer1992eiffel}, Frama-C~\cite{cuoq2012frama}, Java PathFinder~\cite{havelund2000model}, JML~\cite{leavens2006preliminary}, mbeddr~\cite{voelter2012mbeddr}, Spec\#~\cite{barnett2004spec}, Spark Ada~\cite{barnes2012spark}, VeriFast~\cite{jacobs2011verifast}, VCC~\cite{cohen2009vcc}, Whiley~\cite{pearce2013whiley}, and Why3~\cite{filliatre2013why3}. \new{Although KeY~\cite{ahrendt2016deductive} was considered a VA language, it was later removed as it introduced significant noise due to the name’s generality, including, e.g., results from security.}

\subsection{Literature Review}
To understand from existing work what challenges are faced by practitioners when using VA languages, we performed a literature review. Next, we present all the steps followed.

We queried the \href{https://dl.acm.org/}{ACM Digital Library}, \href{https://ieeexplore.ieee.org/Xplore/home.jsp}{IEEE Xplore}, \href{https://www.scopus.com/}{Scopus}, and \href{https://www.webofscience.com/}{Web of Science}. The research query was developed following an iterative methodology. Firstly, we identified the following relevant articles that we would like to find with our research query: \cite{White2017, Todorov201884, Nyberg2018139, schoolderman_is_2019, Hahnle2019345}. Then, we developed an initial search query based on keywords from the collected articles. The resulting query was then tailored to find as many articles as possible, while also containing the previously identified ones. We only accepted articles with ten or fewer years because, considering we are trying to understand developer practices and challenges, older articles could contain issues already addressed or outdated. We also included in the query the set of VA languages identified in \Cref{subsec:VAL_considered}. The resulting query is as follows: \vskip 0.5em
\begin{quote}
\scriptsize\ttfamily
\begin{tabbing}
\textbf{(}\= \,"theorem proving" OR "theorem proof" \\  
\> OR "deductive verification" OR "deductive methods"\\ 
\> OR "formal verification" OR "formal specification"\\
\> OR "contract programming" OR "design by contract"\\
\> OR "programming by contract" OR "Bandera" OR "Dafny"\\
\> OR "Escher C Verifier" OR "Eiffel" OR "Frama-C"\\
\> OR "Java PathFinder" OR "JML" OR "mbeddr" OR "Spec\#"\\ 
\> OR "Spark Ada" OR "VeriFast" OR "VCC" OR "Whiley"\\
\> OR "Why3"\,\textbf{)}\\
AND industr* \\
\=AND \textbf{(}\= \,challeng* OR issue* OR drawback? OR constraint?\\ 
\> \> OR limitation? OR obstacle? OR difficult*\,\textbf{)}\\
\end{tabbing}
\end{quote}

The query only searches in the field \textit{Abstract} because it is the field that all databases support. The queries were conducted %
on 26/12/2023, resulting in 1024 articles. After 729 duplicates were removed, 600 articles were screened.

The following exclusion criteria were applied. These aim to include the maximum number of relevant articles. When in doubt, the full-text screening was applied. 
\emph{Title Screening:} 1) the title is not related to formal verification (or derived terms) \,---\, e.g. ``Solidification Principle in Large Vertical Steel Casting Under the EMS Effect''; 2)
the title mentions formal verification tools that are not VA languages \,---\, e.g. Alloy, SysML.
\emph{Abstract Screening:} 1) the term formal methods or related is used in contexts not related to our research \,---\, e.g. formal verification of startup models; 2) the tools or formal verification methods mentioned are related to the area of software engineering, but the application area is not aligned with the our research \,---\, formal verification of system models.
\emph{Full-Text Screening:} 1) the article does not mention any challenges; 2) the challenges are not related to the application of VA languages (e.g., the article talks about the challenges of integrating Dafny with Solidity).

After this process, nine of the 600 articles were considered relevant and were analyzed (see \ref{subsec:LR_RQ1})\new{, highlighting a significant research gap and reinforcing the relevance of this study.}

\subsection{Stack Overflow Q\&A Study}
This section details how we
extract challenges from 
developer
posts on Stack Overflow,
a widely used Q\&A platform where developers from diverse backgrounds openly discuss real-world programming problems at scale, making it a rich and authentic source for empirical studies~\cite{barua_what_2014,bagherzadeh_going_2019,han_what_2020,abdellatif_challenges_2020,
alamin_low_code_2021,mohamed_quantum_2021,peruma_refactor_2022}.  
However, manually analyzing the vast amounts of data available in these sources may become infeasible, particularly as the dataset size increases.
Therefore, we use topic modeling, 
a popular analytical tool for evaluating data,
as an approach 
to identify groups of similar words, or topics, within the text~\cite{vayansky2020review}.  
One of the most popular algorithms for topic modeling is \emph{Latent Dirichlet Allocation} (LDA)~\cite{blei_latent_2003}, an unsupervised machine learning algorithm that models automatically extracted topics in corpora. It is based on the idea that each document is a mixture of topics, and each topic is a mixture of words. LDA aims to discover how much of each topic is in each document and how much of each word belongs to a particular topic. 
The algorithm outputs probability distributions that quantify the proportion of each topic within a document and the association of each word with a given topic~\cite{silge_text_mining_2024}.
Several previous studies have used unsupervised machine learning, in particular LDA, for categorizing forum posts \cite{barua_what_2014, bagherzadeh_going_2019, han_what_2020, abdellatif_challenges_2020, %
peruma_refactor_2022, alamin_low_code_2021, mohamed_quantum_2021}.

Our forum post analysis with topic modeling follows prior work and involves four steps: a) collecting Stack Overflow data on VA languages, b) pre-processing (cleaning, stop-word removal, stemming/lemmatization), c) training an LDA model, and d) labeling tags based on associated words and topics. %

\paragraph{\bf{Data Collection}}\label{sec:methodology_dataset}
Three popular Stack Overflow post datasets are: SO official data dumps~\cite{stackoverflow_dump_location}, SOTorrent~\cite{baltes_sotorrent_2018}, and SE Data Explorer.\footnote{\url{https://data.stackexchange.com/}} The first is distributed as a very large XML file, which we found difficult to process; the second is not updated since 2020; and the third is available through a human-centered interface, making it difficult to automate data collection. Therefore, %
we used the official Stack Exchange API\footnote{https://api.stackexchange.com/} as, similarly to SE Data Explorer, it provides the newest data at the time of querying, but in a computer-friendly format.

We identified the tags we would use to query the API endpoint and settled with a variation of a list of VA languages presented in \Cref{subsec:VAL_considered}. From our observations of the Stack Overflow tag list,\footnote{\url{https://stackoverflow.com/tags}} only hyphenated tags are supported, so ``Escher C Verifier'', ``Java PathFinder'', and ``Spark Ada'' had to be transformed to accommodate this requirement.  %
In the final mapping between %
VA language names and the tag names, ``Escher C Verifier'', ``Java PathFinder'', and ``Spark Ada'' became ``escher'', ``pathfinder'', and ``spark-ada'', respectively.

Moreover, we decided only to query the \emph{/questions} endpoint because we are looking for challenges commonly found in questions but not in the answers and comments.

We ended up with 1420 questions, most of which were about Dafny and Frama-C. The distribution of questions was Dafny: 530, Frama C: 474, Eiffel: 289, Spark Ada: 55, Why3: 52, JML: 44, and Spec\#: 16.

\paragraph{\bf{Data Pre-processing}}\label{sec:model_training}
Since we gathered data from the Stack Exchange API, the question bodies contain code snippets and tags (similar to \cite{stackoverflow_dump_location} and \cite{baltes_sotorrent_2018}). 
With that in mind, we built four pre-processing pipelines, with each pipeline combining a subset of
the following nine steps.
\vskip 0.1em
\noindent\noindent
{\bf Pre-processing steps.} Based on prior research, we considered the following nine pre-processing steps:

\begin{itemize}
\item \emph{\_\_remove\_code:} %
Removes all tags and code snippets from the body of a given question, wrapping it into a \emph{div} tag, and then performing a depth-first search into its children, removing all \emph{code} and \emph{pre} tags. Only the inner text of the other tags, such as \emph{anchor}, \emph{strong}, \emph{paragraph}, etc, and text that is directly a child of a given tag is kept. The resulting text is then trimmed and joined with a white space. Based on \cite{barua_what_2014}, \cite{bagherzadeh_going_2019}, \cite{han_what_2020}, %
\cite{alamin_low_code_2021}, \cite{mohamed_quantum_2021}, and \cite{peruma_refactor_2022}. %

\item \emph{\_\_gensim\_preprocessing:} %
Performs common pre-processing tasks using the functions \emph{strip\_numeric}, \emph{strip\_multiple\_whitespaces}, and \emph{strip\_punctuation} from the Gensim\footnote{\url{https://radimrehurek.com/gensim/}} library. This step removes digits from the initial string and multiple consecutive white spaces, leaving just one separating the words. Finally, it strips all the punctuation from the input string. Based on \cite{bagherzadeh_going_2019}, \cite{han_what_2020}, %
\cite{alamin_low_code_2021}, \cite{mohamed_quantum_2021}, and \cite{peruma_refactor_2022}.

\item \emph{\_\_lemmatize\_body:} %
Starts by tokenizing the input string using NLTK's \emph{word\_tokenize} function. After tokenizing, we apply stemming using NLTK's WordNet lemmatizer. This lemmatizer needs to receive the part of speech of each word. Since identifying specific parts of speech is out of the scope of our work, we decided to consider every word as a verb since we found that verbs and plural nouns were transformed into their stems. 
Based on \cite{han_what_2020}. 

\item \emph{\_\_remove\_stop\_words:} %
Filters out stop words using the NLTK stop words corpus (e.g., \emph{a}, \emph{an}, \emph{the}, \emph{of}, \emph{and}). This is an approach used by \cite{han_what_2020}, \cite{abdellatif_challenges_2020}, and \cite{peruma_refactor_2022}.

\item \emph{\_\_transform\_bigrams:} Takes a list of words and returns a list of bigrams using NLTK \emph{bigrams} function. This step was also used by \cite{abdellatif_challenges_2020}, although the authors used Gensim to build the bigrams.

\item \emph{\_\_make\_lower\_case:} Transforms all words into their lowercase version.

\item \emph{\_\_remove\_handpicked:} Removes words that, after examination, we considered as not adding value to the trained model \new{\,---\, e.g., \emph{also}, \emph{always}, \emph{etc}, \emph{like} (full list available in the  \href{https://figshare.com/s/bc220b6c300b46c97d89}{supplementary material}).}
Based on \cite{alamin_low_code_2021} and \cite{peruma_refactor_2022}.

\item \emph{\_\_remove\_uncommon:} Removes words with less than three uses since these words can not form patterns. %

\item \emph{\_\_remove\_small:} Removes words with less than two characters since we considered that words with that size could not express meaningful value.

\end{itemize}

\noindent\noindent
{\bf Pre-processing pipelines.} After collecting the data, we process the questions using the four pre-processing pipelines listed below, developed as a combination of the nine steps listed previously. 
All the pipelines are based on \cite{han_what_2020} because they lemmatize the words before removing the stop words. %
\begin{enumerate}
    \item \emph{baseline:} This pipeline is the simplest and produces a list of individual words. The exact sequence of the steps is: \emph{\_\_remove\_code} \(\!\rightarrow\!\) ~\emph{\_\_gensim\_preprocessing} ~\(\!\rightarrow\!\)~\emph{\_\_lemmatize\_body} \(\!\rightarrow\!\) 
 \emph{\_\_make\_lower\_case}  ~\(\!\rightarrow\!\) \emph{\_\_remove\_stop\_words}.
    
    \item \emph{baseline\_bigram:} Create bigrams from the list of words resulting from the baseline pipeline. Based on the work by \cite{abdellatif_challenges_2020}. The steps are: \emph{baseline} \(\!\rightarrow\!\)\emph{\_\_transform\_bigrams}.  
    
    \item \emph{rem\_words\_baseline}: Similar to the regular baseline pipeline but incorporates the innovation of \cite{alamin_low_code_2021} and \cite{peruma_refactor_2022} of removing domain-specific words as well as an additional removal of words with less than three uses\,---\,since these words can not possibly form patterns\,---\,and less than two characters\,---\,since we considered that words with that size could not express meaningful value. The sequence is: \emph{baseline}  \(\!\rightarrow\!\) \emph{\_\_remove\_handpicked}  \(\!\rightarrow\!\)\emph{\_\_remove\_uncommon} \(\!\rightarrow\!\) \emph{\_\_remove\_small}.
    
    \item \emph{rem\_words\_bigram:} Based on the rem\_words\_baseline but, similarly to baseline\_bigram, ends with a step that transforms the resulting list of words in a list of bigrams. Steps: \emph{rem\_words\_baseline} \(\!\rightarrow\!\)\emph{\_\_transform\_bigrams}.

\end{enumerate}

\paragraph{\bf{Model Training}}\label{sec:model_training}
To configure the LDA algorithm, several parameters, known as hyperparameters, must be set: the number of topics ($K$); the number of iterations (i.e., optimization steps); the number of passes over the training corpus; to what extent questions mix topics ($\alpha$)\,---\,a high value means that questions cover a wide range of topics; and to what extent questions mix words ($\beta$)\,---\,a high value means that topics are likely to contain a broad range of words, rather than being focused on a small, specific set of terms.

Each of the four pre-processing pipelines presented 
in the previous subsection
was tested against each combination of the following hyperparameters: \(K\) ranging from 2 to 40 in steps of 2 to capture all identified challenges, and from 500 to 3000 in steps of 500 to align with prior work; we also tried 50 iterations, the Gensim %
default. We set the number of passes at 1 (Gensim default) and 5 (as per \cite{han_what_2020}). For \(\beta\), we tested 0.1, 0.01, and None (Gensim default), while \(\alpha\) followed the common formulation \(\alpha = 50/K\) from prior studies~\cite{bagherzadeh_going_2019, alamin_low_code_2021}.

\vskip 0.1em
\noindent\noindent
{\bf Final Model.}
We select the final model configuration through an evaluation across multiple training runs and criteria, which, due to space limitations, we briefly detail below. More details can be found in the \href{https://figshare.com/s/bc220b6c300b46c97d89}{supplementary material}.\footnote{\new{\url{https://figshare.com/s/bc220b6c300b46c97d89}}}

This process began with an \emph{initial training run} using the full dataset of collected Stack Overflow questions. This run aimed to evaluate the performance of different pre-processing pipelines and model configurations. This produced: 4 model runs, one for each pipeline; 840 model configurations resulting from the combination of LDA hyperparameters; and 3360 models, four for each configuration. To account for the non-deterministic nature of the LDA algorithm, we trained four models for each combination of the hyperparameters. %

Model performance was evaluated using two primary metrics: topic coherence (\(C_{UMass}\))~\cite{roder_exploring_2015} and perplexity score (logarithm of perplexity)~\cite{blei_latent_2003}. %
We decided to use \(C_{UMass}\) to evaluate our models since it is a standard metric that can be quickly calculated. We also analyzed the logarithm of the perplexity, which measures how likely an outside observer is to predict the values drawn from a model. The values used are the output of the function log\_perplexity\footnote{\url{https://radimrehurek.com/gensim/models/ldamodel.html\#gensim.models.ldamodel.LdaModel.log_perplexity} [last access 31st August 2024]} in Gensim. %

We trained a model for each combination of pre-processing pipelines and hyperparameters. The performance of each combination (pipeline $\times$ combination of hyperparameters $\times$ 4 models) was assessed by calculating the average coherence and perplexity scores, as well as their standard deviations across the four models. This allowed to identify configurations that were not only high-performing but also consistent. Configurations with high standard deviations were considered unstable.
Taking into consideration LDA's non-deterministic nature, this strategy allows us to detect if a score is exceptionally good or if the hyperparameters are able to make LDA consistently output good models.

The \emph{baseline} and \emph{baseline\_bigram} pipelines resulted in configurations with significantly high standard deviations in coherence, indicating instability. 
The \emph{rem\_words\_baseline} pipeline showed better overall performance and, unlike \emph{baseline}, it removed words that were not helpful for identifying challenges. 
The \emph{rem\_words\_bigram} pipeline generally showed higher degradation in coherence compared to \emph{rem\_words\_baseline}.
Based on this initial analysis, the \emph{rem\_words\_baseline} was chosen. A configuration with 16 topics within this pipeline was initially selected for further consideration.

As further explained in \ref{sec:topic_labelling}, questions with low probability of belonging to a topic were discarded and open card sorting \cite{spencer_card_2009} was performed based on the top 30 words and top 20 documents for each topic. Looking at the composition of the topics, we obtained 209 unique questions from the 320 that were assigned to the 16 topics. Most questions appeared only once, indicating diverse topics.

\begin{figure}
    \centering
        \caption{Frequency of each tag (initial training)}
        \label{fig:tag_dist}
        \includegraphics[width=0.4\textwidth]{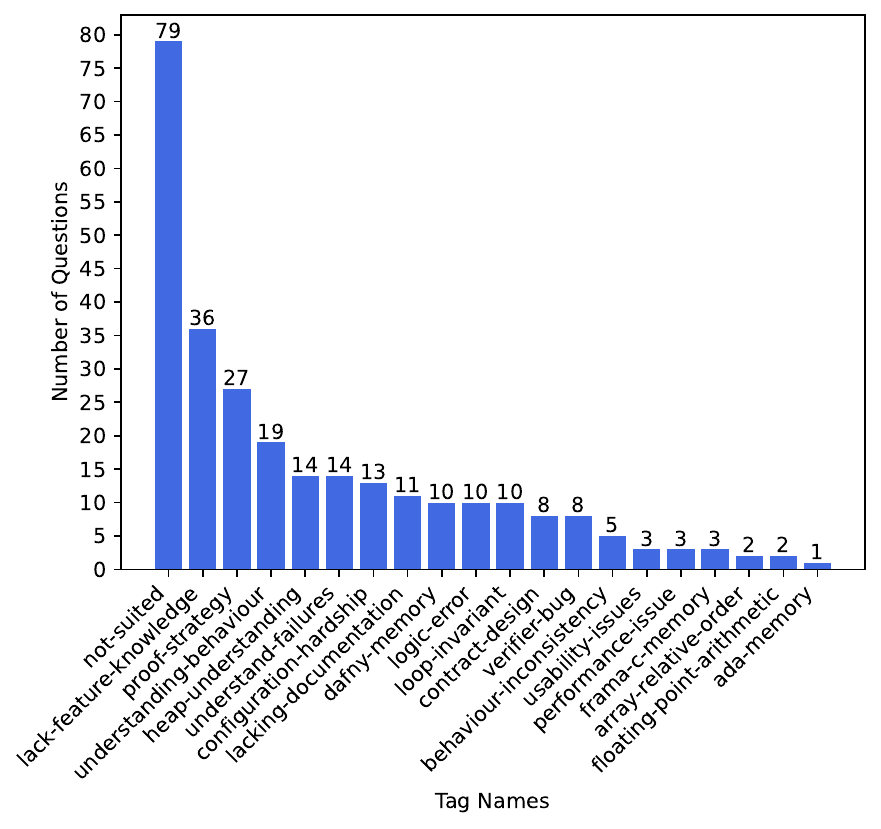}
\end{figure}

After applying open card sorting to the clustered questions, we obtained the tag distribution depicted in \Cref{fig:tag_dist}. The most frequent tag is the \emph{not-suited} tag with 79 questions, meaning that around 38\% of questions were considered inadequate to answer RQ1. Of the 16 resulting topics, in 15, the most frequent tag was \emph{not-suited}, indicating that the topics are not representative of any challenge. %

To overcome this issue, we decided to filter out these questions and run the training in a dataset without the 79 inadequate questions. 
A \emph{second training run}  was then done using the filtered dataset. Again, we applied the four pre-processing pipelines and explored various hyperparameter configurations and used the same evaluation metrics as in the \emph{first training run}. %
Similar performance patterns to those obtained in the \emph{first training run} were observed across the pipelines, with \emph{baseline}, \emph{baseline\_bigram}, and \emph{rem\_words\_bigram} showing high errors.
The \emph{rem\_words\_baseline} pipeline again demonstrated consistent performance for configurations filtered by lowest error.
Considering the similar performance, we again chose the \emph{rem\_words\_baseline} pipeline to pick the best configuration and model. We chose the configuration with eight topics because it resulted in a number of topics that allowed for the identification of challenges consistent with prior literature, as discussed in \Cref{subsec:LR_RQ1}.
Looking at the topics, we observe these are composed of 160 questions, from which 145 are unique. %
This indicates that this model is diverse.

\begin{figure}
    \centering
        \caption{Frequency of each tag (second training)}
        \label{fig:second_tag_dist}
        \includegraphics[width=0.38\textwidth]{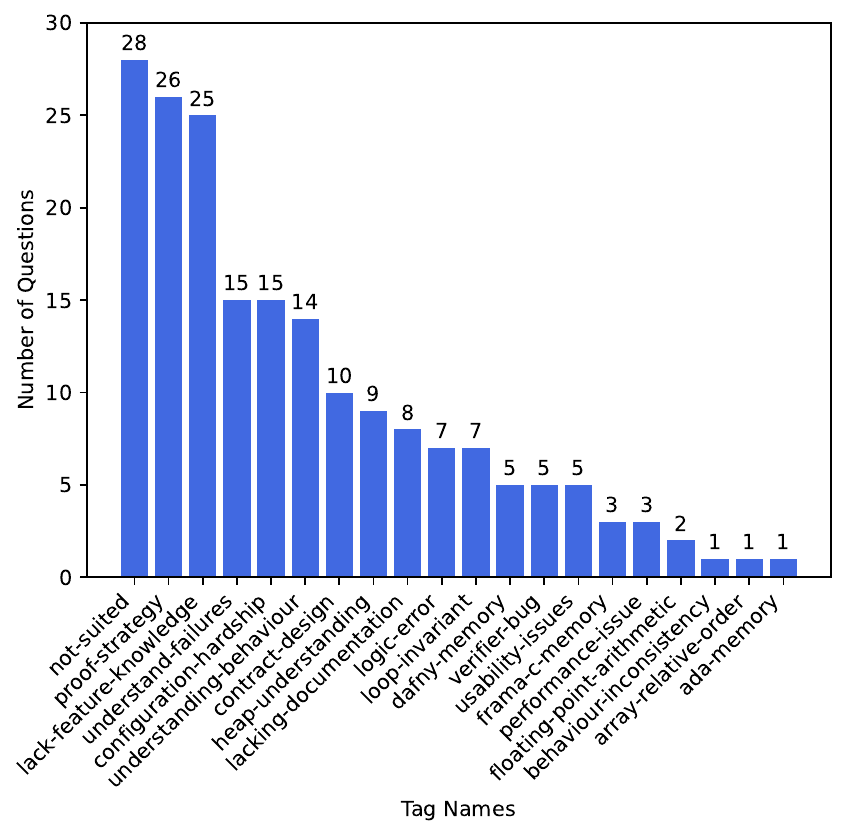}
\end{figure}

After performing open card sorting, we end up with the tag distribution shown in Figure \ref{fig:second_tag_dist}. Despite being the most popular tag, the \emph{not-suited} questions only comprise 19\% of the overall questions against the 38\% obtained in the \emph{initial training run}. %
Based on our analysis, we decided that this model was suited for naming and challenge identification.

\paragraph{\bf{Topic Labelling}}\label{sec:topic_labelling}
This process involved analyzing the chosen model configuration's output. Questions %
were filtered, discarding those where the probability of belonging to a topic was less than \(\frac{1}{n}\) (where $n$ is the number of topics). The reasoning for this was that \(\frac{1}{n}\) is the probability of randomly assigning a document to a given topic, thus we only accepted probabilities greater than the random assignment.

After filtering the questions with low probabilities, we sorted them and performed open card sorting \cite{spencer_card_2009} based on the top 30 words and top 20 documents for each topic. The labeling for each topic was discussed among the team members until a consensus was reached.

For the final chosen model configuration, which had eight topics, the labeling process involved analyzing the frequency and distribution of the tags that were created during the open card sorting process.

For \emph{Topic 0}, even though the most common tag is \emph{not-suited}, the label {\bf Memory Model Understanding} was assigned, influenced by the second most common tag, \emph{heap-understanding}.
\emph{Topic 1} was labeled {\bf Language Understanding}, driven by the frequency of the \emph{lack-feature-knowledge} and \emph{understand-behaviour} tags, and the occurrences of \emph{logic-error}, \emph{loop-invariant}, \emph{understand-failures}, and \emph{lacking-documentation}. 
\emph{Topic 2} was labeled {\bf Not Suited} because no clear relationship between the clustered questions was found.
\emph{Topic 3} was labeled {\bf Environment Setup} as most of the questions are related to to the \emph{configuration-hardship} tag.
\emph{Topics 4} and \emph{5} were labeled {\bf Non Relevant} as no clear relationships were found between the top question label.
\emph{Topics 6} and \emph{7} were labeled \textbf{Proof Elaboration}, because both contain many questions regarding the label \emph{proof-strategy} and some questions regarding \emph{loop-invariant}.
These labeled topics were then used to identify broader categories of challenges, which are presented in the \Cref{sec:results_SO}.

\subsection{User Study}
To better understand the challenges faced by VA language users and assess the relevance of the issues identified in our prior qualitative analysis, we conducted an online survey study. %
We designed the study to capture both quantitative and qualitative feedback, drawing on the experiences of participants with different levels of familiarity with VA languages. %
We provide the full protocol in the \href{https://figshare.com/s/bc220b6c300b46c97d89}{supplementary material}.

\subsubsection{Survey Design}
We start the survey with a section about background information, where we ask about familiarity with VA languages and years of experience. We also ask about participants' initial contact with VA languages and how they learned to program with them.
We then attempt to validate the answers to RQ1. We use matrix-style questions with a 5-point Likert scale to ask questions about challenges related to \textbf{Proof Elaboration} and \textbf{Environment Setup}. 
We also ask participants about a subset of \textbf{language-agnostic challenges} identified in \Cref{sec:results_SO} and \textbf{language-specific challenges} with Dafny and Frama-C. We only showed these sections to participants who mentioned having experience with these languages. \new{This part of the survey mainly targets Dafny and Frama-C practitioners because over 70\% of the questions found in Stack Overflow were related to these specific languages and some challenges are language dependent.} 

We ask two open-ended questions about additional challenges not identified in the previous parts of the study and suggestions for improving the adoption of VA languages. Then, we use a matrix-style closed-ended question to ask about the suggestions we developed \textit{a priori}. %
We asked open-ended questions before the closed-ended to prevent bias and allow the participants to think
more freely without being conditioned by the options they had already seen.

\subsubsection{Recruitment}
\label{sec:validation_survey_distribution}

Our sample had a total of 31 participants. We recruited participants using our social media, professional network, and snowball sampling. We began by distributing the survey to 15 students of the 2023/2024 edition of %
the Formal Methods for Critical Systems course, an optional first-year Master's level course from the Faculty of Engineering, University of Porto. %
Using Proxycurl,\footnote{\url{https://nubela.co/proxycurl/}}  
we gathered 28 profiles of LinkedIn users from the US with software and formal verification skills, contacted 16 industry professionals from our professional network,  %
and 26 Dafny contributors on GitHub, whose contact information was public. Additionally, we shared our survey on social media through \emph{X}.\footnote{\url{https://x.com/}} We used snowball sampling by asking our contacts to distribute the survey to relevant members of their professional and student networks.
We also distributed the survey through Frama-C's mailing list. 
We divide participants by years of experience: more than three are considered \textit{experienced}, all others are \textit{beginners}.

\new{All participants were over 18 years old and provided informed consent after reviewing a form with the study’s purpose, procedures, and their rights. Participants were informed that they were not required to participate and could withdraw at any time without penalty. No personal or identifiable data were collected.}

\textbf{Sample.}  
Most of our sample identify as male and are employed in the computing field with formal education. Most experienced participants have some graduate or professional degree, while most beginners only have a bachelor's degree.
 See \Cref{tab:demographics} for complete demographic information.

\begin{table}
 \centering
 \small

 \begin{table}[H]
\centering
\caption{Overview of participant demographics and experience}
\label{tab:demographics}
\begin{tabular}{@{}lrrr@{}}
\toprule
& \textbf{Beginners} & \textbf{Experienced} & \textbf{Total (\%)} \\
\midrule
\textit{Gender} \\
Male                  & 14 & 11 & 80.6\% \\
Female                &  4 &  1 & 16.1\% \\
Prefer not to say     & -- &  1 & 3.2\% \\

\midrule
\textit{Age} \\
18--24 years          & 15 &  1 & 51.6\% \\
25--34 years          &  3 &  2 & 16.1\% \\
35--44 years          & -- &  7 & 22.6\% \\
45--64 years          & -- &  3 & 9.7\% \\

\midrule

\textit{Highest Education Achieved} \\
Some college, no degree         & 2  & -- & 6.5\% \\
Associate's or technical degree & -- & 1  & 3.2\% \\
Bachelor's degree               & 14 & 1  & 48.4\% \\
Graduate/professional degree    & 2  & 11 & 41.9\% \\

\midrule
\textit{Race} \\
Asian                & 1  & 1  & 6.5\% \\
Black                & 4  & -- & 12.9\% \\
White                & 12 & 8  & 64.5\% \\
Prefer not to say    & 1  & 4  & 16.1\% \\

\midrule
\textit{Currently Employed in CS} \\
Yes                           & 7  & 13 & 64.5\% \\
No                            & 9  & -- & 29.0\% \\
No, but have been in the past & 2  & -- & 6.5\% \\

\midrule
\textit{Formal CS Education} \\
Yes & 18 & 11 & 93.5\% \\
No  & -- &  2 & 6.5\% \\

\addlinespace
\textbf{Total Participants} & \textbf{18} & \textbf{13} & \textbf{100\%} \\
\bottomrule
\end{tabular}
\end{table}

\end{table}

\subsubsection{Quantitative Analysis}

For our quantitative data, we use the \emph{Mann-Whitney U} test to check whether the data from two groups, beginners and experienced, differ significantly. We chose a non-parametric test because our data does not follow a normal distribution. 
We use a Friedman test to compare answers to statements within the same group. If the Friedman test is significant ($p < 0.05$), we then use the Wilcoxon signed-rank test to make pairwise comparisons of each statement and understand which particular statements are significantly different from the others.  
Finally, we analyze the frequency of each response to understand %
if the participants encountered the challenge. %
To analyze Likert scale data, we remove the \emph{Not Sure} responses from the statistical test, assuming that these were due to the participant not understanding the statement. %

\subsubsection{Qualitative Analysis}

Most of our survey consisted of closed-ended questions. However, we asked two open-ended questions about challenges and possible solutions to challenges. We used deductive and inductive coding to analyze the qualitative data derived from these questions. 
Two coders did two independent rounds of coding. In the first round, the initial coder designed the first version of the codebooks that both coders used. This first codebook was developed with prior knowledge of the types of challenges identified in an earlier portion of the study. After completing the first pass, the codebook was refined, and the coders met to discuss the added and modified codes. Following these updates, both coders independently applied the revised codebooks to the dataset in a second round of coding. After this, the coders met to resolve discrepancies in their coding and reached a consensus. The process resulted in two finalized codebooks: one for suggested improvements, consisting of 13 codes, and another for challenges, consisting of 14 codes. Each participant's response could have up to 5 different codes. %

\section{What challenges do Verification-Aware language practitioners face? (RQ1)}

\subsection{Findings from Literature Review}\label{subsec:LR_RQ1}
From the existing literature, we identified several challenges, such as: %
\emph{behavior dependent on previous executions}, which include challenges related to the verification of properties where the behavior is influenced by previous executions~\cite{dordowsky_experimental_2015, gurov_deductive_2017}; %
\emph{behaviour inconsistency}, as verification success changes depending on the solver~\cite{Todorov201884}; \emph{cost of applying VA languages}~\cite{White2017}; \emph{feedback quality}, which includes the lack of counter-examples by some tools~\cite{Todorov201884}, the lack of clarity of the error messages~\cite{schoolderman_is_2019}, and a general difficulty in understanding why proofs fail~\cite{gudemann_online_2021}; \emph{handling boolean values}, i.e. challenges related to verifying properties about boolean values~\cite{gurov_deductive_2017, djoudi_formal_2021}; \emph{imprecise specifications}~\cite{gurov_deductive_2017, knuppel_experience_2018}; %
\emph{increasing the scale of verification}, which includes challenges related to the integration of multiple verified functions~\cite{knuppel_experience_2018}, the scaling of verification time~\cite{djoudi_formal_2021}, the difficulty of tracking and maintaining annotations~\cite{djoudi_formal_2021, Hahnle2019345}, and general claims about scalability~\cite{Nyberg2018139}; \emph{integration with hardware}, including difficulties related to the verification of functions that interact with hardware, often resulting in side effects~\cite{dordowsky_experimental_2015}; \emph{integration with other tools}~\cite{Hahnle2019345}; \emph{lack of features} to support verifying some behaviors~\cite{knuppel_experience_2018, selvaraj_verification_2019}; \emph{lack of verified APIs}~\cite{Hahnle2019345}; %
\emph{lacking support for specific tasks}, such as lack of support for floating-point arithmetic~\cite{Nyberg2018139, Hahnle2019345}, concurrent software~\cite{Hahnle2019345}, and reflection~\cite{Hahnle2019345} verification; \emph{loop invariant discovery}~\cite{schoolderman_is_2019, gudemann_online_2021}; \emph{need for automation}, which includes the need for automation so that deductive verification can be widespread~\cite{Nyberg2018139}; %
\emph{insufficient~\cite{Todorov201884, schoolderman_is_2019, Hahnle2019345} and costly\cite{White2017, Hahnle2019345} support and documentation}; 
\emph{switching costs}, which includes authors that considered the lack of interoperability between tools and the cost to switch them an obstacle for the adoption of VA languages~\cite{White2017}; and \emph{type conversions}, including challenges related to the verification of implicit~\cite{gurov_deductive_2017} or heterogeneous~\cite{djoudi_formal_2021} type conversions.

\subsection{Findings from Stack Overflow Q\&A Study} \label{sec:results_SO}
In this section, we group the identified challenges by the four topics identified in \ref{sec:topic_labelling}. Due to space restrictions, full details can be found in the \href{https://figshare.com/s/bc220b6c300b46c97d89}{supplementary material}.

\paragraph{\bf Proof Elaboration} \label{sec:results_challenges_proof_elaboration}
The challenges in this class are related to the programming paradigm, so these are language agnostic. %
The first challenge is related to creating a proof strategy. In many questions, authors could not even start the proof, not knowing what proof strategy to choose. This challenge was observed in nine questions.

The second challenge is related to completing a proof. In contrast with the previous challenge, these authors could start to design a proof strategy that then proved impractical in verifying the properties of the code. This challenge was observed in three questions. %

The third challenge is related to authors who chose a non-ideal strategy to prove the properties of the code; for example, choosing a proof by induction when a proof by contradiction would be easier. This was observed in one question. 

The fourth challenge concerns identifying and expressing the necessary loop invariants to verify loops. This challenge was observed in five questions. %

The fifth challenge, observed in one question, is related to the fourth and concerns identifying and expressing a valid decrease clause that proves that loops or recursion functions terminate. This challenge was observed in one question. %

The sixth challenge concerns not being able to gather feedback, i.e., whether the contracts are respected due to the verifier timing out. This was observed in two questions.  %

\paragraph{\bf  Environment Setup} \label{sec:results_challenges_env_setup} 
The challenges in this topic are language agnostic.
The first class of challenges is related to dependency installation, which were identified in four questions. It includes challenges related to missing dependencies, incompatible dependency versions, and dependencies not being found. %
The other two classes of challenges are related to unsupported use cases and missing installation instructions. These challenges are represented by one question each. %

\paragraph{\bf Memory Model Understanding} The challenges in this topic are language-specific, \new{as we observed that languages where the user manually manipulates the memory, like C, have different challenges than garbage-collected ones, like Dafny. }%

We identified three challenges among Dafny programmers, all related to the \emph{modifies} clause. The \emph{modifies} clause specifies the set of memory locations that a method, iterator, or loop body may modify~\cite{dafny_reference_manual}.

The first challenge is related to the user applying a \emph{modifies} clause to a given memory location, holding a reference to an object within the scope of a method, but only changing the properties of the given object inside a loop. Dafny assumes that the loop might change the object to which the memory location is pointing, thus making verification based on the previously set object properties impossible. The programmer must create an invariant saying that the memory location always points to the same object.

The second challenge is related to programmers not understanding which variables are affected by the \emph{modifies} clause, identified in two questions. %
In both, we see that the authors fail to acknowledge what is the right target of the \emph{modifies} clause when the target property is a memory location to an object: one author is missing a \emph{modifies this} clause; the other tries to modify a property of the class while using \emph{modifies this} instead of including \emph{modifies this.property\_name}.

In another question, we observed a similar challenge, but the clause the author is using is the \emph{reads} clause. The \emph{reads} clause marks all the heap memory locations the function is allowed to read \cite{dafny_reference_manual}. %

\paragraph{\bf  Language Understanding} \label{sec:results_challenges_lang_understand}
Similarly to the previous topic, most of the challenges in this topic are related to features that are language-specific, related to Dafny, and to the Frama-C WP Plugin, which uses ACSL as an annotation language.

For Dafny, in one of the questions, the author 
seemed to lack knowledge about partial functions.
The second question is about the \emph{fuel} attribute in Dafny using an example function.
Another question concerns an author that fails to understand the error message provided by the verifier about a type definition. Upon further inspection, we considered the message self-explanatory for developers who knew about subset types and witnesses. %
In the final question, the author fails to acknowledge what constructs
are opaque. In Dafny, when a construct is opaque, the verifier does not introspect it to derive pre-conditions and post-conditions\,---\,the developer must define the contract. In this specific case, the developer forgot the constructor's post-condition.

Regarding the Frama-C questions, in one of the questions, the author is trying to use a feature defined in the ACSL Reference Manual but not implemented by the Frama-C WP Plugin. In another, the author wants to prove 
that, given the same arguments, the function always returns the same values using ACSL. The accepted answer claims that this is impossible in ACSL. %
In the third question, the author is trying to use a C11 specification feature not implemented by the Frama-C pre-processor at the time of writing.
In the final Frama-C question, the accepted answer claims that the author fails to acknowledge the need for \emph{loop assigns} in the presented loops. The question has some \emph{loop assigns} expressions and was edited 8 hours after creation, presumably to add them. %

The following questions are considered language agnostic. In the first, the author wants to prove a lemma in Dafny. The original code of the author causes an internal error in Dafny 4.7.0. According to the accepted answer, an assertion was missing to make the verifier accept the proof. Because the need for the assertion is probably tied to the internal verification mechanism, we considered that the proof elicits a lack of language understanding, namely, the underlying verification mechanisms. In the second question, the author is trying to understand the translation of mapping WhyML code into the SMT encoding to understand how the proofs for ACSL annotated programs are provided. This question is not a challenge but introduces a need for understanding the underlying verification mechanisms. The last question contains a simple syntax error where the author does not use the correct syntax to write comments in ACSL.

\subsection{Findings from User Study}

\paragraph{\bf Proof Elaboration Challenges}
In this section, we validate seven challenges related to \emph{Proof Elaboration}. %
 Of the seven challenges, presented in \Cref{tab:proof-challenges}, only \(PE_7\), which addresses difficulties in expressing the necessary clause to prove loop/recursion termination, had a significant p-value (\(p = 0.02666\)), this means that \(PE_7\) is the only PE challenge significantly worse for beginners than experienced users. All other challenges 
had a non-significant p-value (\(p > 0.05\)) when comparing beginners and experienced users. 

To assess differences between the seven conditions across all participants, we conducted a Friedman test.
The results indicate a significant difference between the conditions, \(chi^2(6) = 27.79\), \(p = 0.0001\). To assess these differences, we conducted pairwise comparisons with a Wilcoxon signed-rank test with Bonferroni correction (see~\Cref{tab:pairwise_wilcox}). The results indicate that there were significant differences between \(PE_1\) and \(PE_7\) (\(p = 0.023\)) and between \(PE_2\) and \(PE_7\) (\(p = 0.032\)). No other comparisons reached statistical significance after the Bonferroni correction (all \(p > 0.05\)).
This means that \(PE_7\), the challenge of expressing loop/recursion termination clauses, is a significantly less important problem than \(PE_1\)\,---\,designing proof strategies\,---\,and \(PE_2\)\,---\,completing proof strategies.

\begin{table}
\centering
\caption{%
Proof Elaboration Challenges}
\begin{tabular}{ll}
\hline
\textbf{ID} & \textbf{Challenge Description} \\
\hline
\(PE_1\) & Difficulty designing proof strategies accepted by the validator. \\
\(PE_2\) & Difficulty completing designed proof strategies. \\
\(PE_3\) & Tendency to choose non-ideal proof strategies. \\
\(PE_4\) & Inability to identify necessary loop invariants. \\
\(PE_5\) & Inability to express necessary loop invariants. \\
\(PE_6\) & Inability to identify clauses proving termination. \\
\(PE_7\) & Inability to express clauses proving termination. \\
\hline
\end{tabular}
\label{tab:proof-challenges}
\end{table}

\begin{table*}[h!]
\footnotesize
\centering
\caption{Pairwise Wilcoxon signed-rank test results with Bonferroni correction for multiple comparisons. *Significant at adjusted \(p-value < 0.05\).}
\begin{tabular}{lccccccc}
\toprule
\textbf{ } & \textbf{\(PE_1\) } & \textbf{\(PE_2\) } & \textbf{\(PE_3\) } & \textbf{\(PE_4\) } & \textbf{\(PE_5\) } & \textbf{\(PE_6\) } & \textbf{\(PE_7\) } \\
\midrule
\textbf{\(PE_1\) } & - & 95.5, 1.000 & 144.0, 1.000 & 172.5, 0.237 & 150.5, 0.093 & 207.5, 0.172 & 263.5, 0.023* \\
\textbf{\(PE_2\) } & - & - & 80.5, 1.000 & 143.5, 1.000 & 139.0, 0.368 & 169.0, 0.336 & 189.5, 0.032* \\
\textbf{\(PE_3\) } & - & - & - & 178.5, 1.000 & 148.5, 0.611 & 205.5, 0.811 & 184.5, 0.058 \\
\textbf{\(PE_4\) } & - & - & - & - & 26.5, 1.000 & 60.0, 1.000 & 86.0, 1.000 \\
\textbf{\(PE_5\) } & - & - & - & - & - & 104.5, 1.000 & 88.0, 1.000 \\
\textbf{\(PE_6\) } & - & - & - & - & - & - & 49.5, 1.000 \\
\textbf{\(PE_7\) } & - & - & - & - & - & - & - \\
\bottomrule
\end{tabular}
\label{tab:pairwise_wilcox}
\end{table*}

Our results show that most developers struggled with designing proof strategies (\(PE_1\), \(64.5\%\)). 
The challenge related to difficulty completing proof strategies (\(PE_2\)) was more split with \(48.4\%\) agreeing that this is an issue for them and \(22.6\%\) disagreeing.
When it comes to searching for the right proof strategies (\(PE_3\)), we saw a similar split with \(48.4\%\) agreeing and \(22.6\%\) disagreeing it is a challenge. 
Identifying (\(PE_4\)) and expressing (\(PE_5\)) loop invariants and termination clauses (\(PE_6\) and \(PE_7\)) does not seem to be a problem for most. %

\paragraph{\bf Environment Setup Challenges}
While we do not see a significant difference between beginners and experts, %
most developers did not encounter dependency conflicts when installing verification tools %
with \(67.7\%\) agreeing this is not a problem. 
Similarly, the challenge of finding the correct version of a verification tool compatible with a preferred IDE %
was not common (\(77.4\%\) did not consider this a challenge). %

\paragraph{\bf Feature Knowledge Challenges}
We do not see a significant difference among experienced users. However, most developers, particularly beginners, struggled to understand how their high-level code is translated to the underlying SMT code, %
while more experienced developers felt confident in this area. When it came to translating proofs from other mediums (e.g., paper) %
to the VA language, most developers did not report significant difficulties. Experienced developers showed relatively higher agreement with this challenge than beginners. %

\paragraph{\bf Dafny Challenges}
Most participants did not face significant issues understanding the ``modifies'' and ``reads'' clauses or distinguishing between opaque and non-opaque statements. However, some developers still struggled with expressing the immutability of variables, and many participants were unfamiliar with critical concepts like the fuel attribute, partial functions for reasoning about high-order functions, and using witnesses to prove the non-emptiness of type definitions.

\paragraph{\bf Frama-C WP Plugin Challenges}
\label{sec:validation_results_frama_c}

None of these challenges had a significant p-value  (\(p < 0.05\)) when comparing beginners and experienced users.
Most developers felt confident in their understanding of the different memory models and their ability to choose the correct model for proving code properties and in selecting variables for ``loop assigns'' statements. Some developers also identified missing features in Frama-C’s C specifications and noted that certain features available in other plugins were absent in the WP Plugin.

\paragraph{\bf Additional Challenges}

The most frequently reported challenges 
were ``Tool Performance Issues'' and ``Poor Error Messages and Feedback,'' each mentioned seven times.  ``Complex Language Features and Low-Level Operations'' was reported six times, followed by ``Specification and Annotation Challenges'' with five occurrences. Both ``Understanding Verifier Capabilities'' and ``Learning Curve'' were identified four times each. Additional challenges included ``Mindset Shift,'' ``Scalability and Large Project Integration,'' and ``IDE Integration,'' each reported twice. The following challenges were identified once each: ``Lack of Documentation,'' ``Conflicts with Performance Optimization,'' ``Proof Stability,'' ``Translation of Algorithms,'' and ``Lack of Proof Automation.''

\begin{tcolorbox}[
    enhanced jigsaw,
    sharp corners,
    boxrule=0.5pt, 
    colback=black!9!white,
    boxrule=0pt, 
    frame empty
 ]
\textbf{RQ1:}
VA language practitioners most commonly face challenges in designing and completing proof strategies. The only challenge different between beginners and experienced users was proving loop/recursion termination, though this issue was uncommon and significantly less important than the challenges of designing and completing proof strategies. 
Language-specific semantics (e.g., frame specifications, termination clauses) appeared on Stack Overflow but were not confirmed as challenges in our user study. While environment setup is less problematic, feature knowledge challenges were confirmed, with beginners struggling more with SMT translation, and experienced developers with translating proofs to the VA language.
We identified fourteen new challenges, the most reported ones being ``Tool Performance Issues'', ``Poor Error Messages and Feedback'', ``Complex Language Features and Low-Level Operations'', and ``Specification and Annotation Challenges''.

\end{tcolorbox}

\section{What is the context of the first contact with Verification-Aware languages? (RQ2)}

To understand the educational and professional pathways that lead individuals to work with VA languages, we asked all participants two questions, one about where they first \textbf{learned about} VA languages and another about where they \textbf{learned to program} with VA languages. %
In our sample, we have 18 beginners and 13 experienced programmers.
Most beginners had their first contact and learned VA languages in an \textit{optional university course} (16 participants, 89\%), with the exception of two participants who had their first contact in a work context. On the other hand, most experienced developers had contact with VA languages in a \textit{work context} (12 participants, 92\%). 
Overall, the most frequent response for both groups was a university course (77\%) and work context (48\%). Three participants mentioned online discussions, one mentioned ``university and enterprise colleagues'' (P17), another reported ``self-learning'' (P18), and finally, one participant said their first encounter was due to being ``bored in high school'' (P28). 
From our observation, it seems that most participants learn about VA languages as a choice of education or career path. 
Table~\ref{table:first-learn-about-va} summarizes the results.
\begin{table}%
\centering
\caption{Characterization of participants based on the context in which they first learned about verification-aware languages.}
\begin{tabular}{lr}
\toprule
\textbf{Context} & \textbf{Frequency} \\
\midrule
Optional University Course               & 22 \\
Work Context                             & 15 \\
Online Discussions                       & 3  \\
Mandatory University Course              & 2  \\
With university and enterprise colleagues & 1  \\
Self-learning                            & 1  \\
Was bored in high school                 & 1  \\
\bottomrule
\end{tabular}
\label{table:first-learn-about-va}
\end{table}

When it comes to their experience programming with VA languages, their responses were consistent with their first contact. Most participants reported learning how to program in a university course (22 participants, 71\%) and being self-taught (14 participants, 45\%). When considering only experienced users, 11 participants (85\%) are self-taught and 10 participants (77\%) learned in a work context.

\new{These results suggest that beginners usually first encounter VA languages at university, while experienced developers do so in professional settings and increasingly through self-teaching. %
These insights highlight the need for differentiated learning resources: universities should integrate VA languages into curricula for early exposure, while practitioners could benefit from self-guided resources and community support.
}

\begin{tcolorbox}[
    enhanced jigsaw,
    sharp corners,
    boxrule=0.5pt, 
    colback=black!9!white,
    boxrule=0pt, 
    frame empty
 ]
\textbf{RQ2:} The context of the first contact with VA languages seems to vary by experience level. Most beginners first encounter VA language in university courses, while most experienced programmers are introduced to VA languages in a work context. Moreover, as users gain experience, self-teaching emerges as a key mechanism through which they acquire knowledge.
\end{tcolorbox}

\begin{table}%
\centering
\caption{Codebook with code frequency for the proposed solutions suggested by participants}
\begin{tabular}{rl}
\toprule
\textbf{\#} & \textbf{Code} \\
\midrule
5 & Better Documentation and Learning Materials  \\
4 & Integration with Mainstream Languages and Tools  \\
4 & Improved Usability  \\
3 & Better Error Messages and Feedback  \\
3 & Improved Automation  \\
3 & Increased Awareness and Education  \\
3 & Better Debugging and Profiling Tools  \\
3 & Industry Demand and Investment  \\
2 & Easier Installation and Better Tool Support  \\
2 & Use of Common Specification Methodologies  \\
2 & Skepticism about Practical Use  \\
2 & Reusable Blocks  \\
1 & Exposing Internal Proof State  \\
\bottomrule
\end{tabular}
\label{tab:codebook_frequencies_new}
\end{table}
\section{What improvements can be made to existing tools to overcome those challenges? (RQ3)}

The user survey combined closed and open-ended questions.
The closed questions aimed to validate %
the following specific proposals derived from our earlier analysis: 
1) Create user-readable counter-examples to complement error messages. From our analysis, counter-examples do not always seem helpful and their use was even discouraged in some cases; 
2) Develop official tutorials on common verification exercises (e.g., Fibonacci sequence, greatest common divisor). Neither Dafny nor Frama-C provides guided tutorials on their official pages;
3) Implement a search feature in the documentation. Neither Dafny nor Frama-C include a search feature, and Frama-C's documentation is only available in PDF format. Dafny's online documentation lacks a persistent table of contents, unlike popular tools like Jest,\footnote{\href{https://jestjs.io/docs/getting-started}{https://jestjs.io/docs/getting-started} [last access 16th July 2024]} %
and Pytest;\footnote{\href{https://docs.pytest.org/en/8.2.x/contents.html}{https://docs.pytest.org/en/8.2.x/contents.html} [last access 16th July 2024]}
4) Expand the themes covered in the documentation based on the gaps identified in our analysis; and
5) Promote VA languages in mandatory courses, as these languages often appear only in optional courses, unlike testing, which is frequently taught.

The open-ended responses complemented these results with additional, user-driven suggestions not previously captured. In this subsection, we present both the quantitative results and the qualitative themes that emerged from participants’ feedback. %

\vskip 0.1em
\noindent\noindent
{\bf Closed-ended Questions.}
The first improvement addresses the inclusion of user-readable counter-examples that could allow users to ``debug'' the proofs. %
Most participants (83.9\%) agreed on its importance. 
The suggestion to improve tutorials with common projects %
was also well-received by the majority of participants (64.5\%). 
Improving documentation search mechanisms 
saw a notable divide, with beginners (66.7\%) placing more emphasis on this improvement than experienced developers (15.4\%), and this difference was statistically significant (\(p = 0.02822\)). 
Expanding documentation themes 
received mixed responses, with 41.9\% considering it very important. 
Finally, including VA languages in undergraduate programs
was considered highly impactful by most participants (50.1\%).

\vskip 0.1em
\noindent\noindent
{\bf Open-ended Questions.}
We identified 13 categories of improvement for VA languages (see~\Cref{tab:codebook_frequencies_new}). 
The most frequently mentioned area was \textbf{improvements in documentation and learning materials}, suggested by five participants. These requests included better tutorials, guides, and documentation to assist both newcomers and experienced developers in using VA languages effectively. For example, one beginner developer, P28, noted the need for ``more documentation on coming up with invariants''.
Four participants mentioned \textbf{improved usability}, and three better \textbf{education}.
When facing errors, two beginners and one experienced developer mentioned that \textbf{better debugging} would help them understand why the verification fails. 
P23 mentioned directly that ``To be really useful in practice, a verification-aware language needs to be familiar, easy to write proofs in, predictable in terms of what can be verified, well-integrated with common tools, interoperable with other languages, and able to generate efficient code.'' (P23). This aligns with the broader theme of \textbf{integration with mainstream languages}, mentioned by four participants.

Another suggestion by two participants is related to \textbf{reusable blocks}, P19 proposed that tools should ``provide more automation to help `easy tasks', so the programmer/prover can focus on complex tasks'', and  P26, mentioned directly they would like to see ``reusable bricks for common verification patterns``. 
Three participants advocated for increased \textbf{investment in verification tools}, suggesting that funding and institutional support are necessary for widespread adoption.
While the majority of participants maintained a positive outlook, \textbf{skepticism} regarding practical adoption was expressed by two participants. For instance, P13 remarked, ``verification is expensive, and tends not to produce immediate value, something which corporate doesn't like very much.''

\begin{tcolorbox}[
    enhanced jigsaw,
    sharp corners,
    boxrule=0.5pt, 
    colback=black!9!white,
    boxrule=0pt, 
    frame empty
 ]
\textbf{RQ3:} 
Our findings point to four key improvements for addressing challenges in VA languages: better debugging tools; richer educational resources like tutorials and improved documentation; greater automation of routine verification tasks; and tighter integration with mainstream development tools. %
\end{tcolorbox}

\section{Related Work}
To the best of our knowledge, this is the first work to address challenges faced by developers using VA languages. In this section, we focus on related work on analysis and topic modeling of developer discussions, and on studies on challenges faced when using different programming languages.

\vskip 0.1em
\noindent\noindent
{\bf Developer Discussion Analysis and Topic Modeling.}
Prior studies have applied topic modeling %
to analyse developer discussions on platforms like Stack Overflow and GitHub, focusing on areas different from VA languages~\cite{carreira2025worse}.
Barua et al.~\cite{barua_what_2014} analysed posts from a Stack Overflow data dump %
to understand the main general discussion topics on the platform.
Bagherzadeh and Khatchadourian~\cite{bagherzadeh_going_2019} used the official Stack Overflow data dump to study the topics of questions %
about big data. Han et al.~\cite{han_what_2020} studied what issues deep learning framework practitioners discuss, gathering data from Stack Overflow %
and GitHub. %
Abdellatif et al.~\cite{abdellatif_challenges_2020} used the official Stack Overflow data dump %
to identify the topics chatbot developers were asking about. They analysed heuristics to discover relevant tags, %
using only post titles for training data, claiming it was representative and reduced noise.
El aoun et al.~\cite{mohamed_quantum_2021} investigated challenges in quantum software engineering using data %
from GitHub and the Stack Exchange Data Explorer (for Stack Overflow).
Al Alamin et al.~\cite{alamin_low_code_2021} conducted an empirical study of developer discussions on low-code software development challenges. 
Peruma et al.~\cite{peruma_refactor_2022} used the SOTorrent dataset to study the topics around Software Refactoring being asked by developers. %
These studies %
demonstrate a well-established methodology for using topic modeling, primarily on online forum data, to study %
developer discussions across various software engineering areas. Our work builds on these studies %
and applies them to the specific context of VA languages.

\vskip 0.1em
\noindent\noindent
{\bf Programming Language Challenges.}
Other studies have examined the challenges developers face when using programming languages, suggesting that even experienced developers struggle to learn new programming languages~\cite{shrestha2020here}. Ko et al.~\cite{ko2004six} identified six learning barriers in programming systems, which include understanding, design, use, and information barriers. These barriers align closely with the challenges reported by VA language users in our study, such as difficulty interpreting tool feedback and expressing proof strategies. 
Other studies focused on particular languages. For instance, Zhu et al.~\cite{zhu2022learning} focused on Rust. %
They analyzed 15,000 Stack Overflow posts and surveyed 101 Rust developers to find that Rust safety features impose a steep learning curve and that compiler messages often lack actionable information. This is in line with other work that identified onboarding and usability issues as barriers to the adoption of languages~\cite {fulton2021benefits,johnson2013don}.
Chakraborty et al.~\cite{chakraborty2021developers} examined the support ecosystem for newer languages such as Go, Swift, and Rust by analyzing both Stack Overflow and GitHub data. They identified gaps in documentation, which often drove developers to rely on community Q\&A sites. 
Finally, Christakis and Bird~\cite{christakis2016developers} studied developer expectations regarding program analysis tools by surveying engineers at Microsoft. They found that high false-positive rates, poor tool integration, and unclear feedback were the main reasons developers rejected static analysis tools in practice. 
These findings echo the feedback-related problems reported by VA language users in our study and suggest that if tools fail to align with developers' workflows or generate opaque error messages, they are unlikely to be adopted.

\section{Conclusion}

This work analyzes the challenges developers face with VA languages, incorporating feedback from both beginners and experienced users. We trained an ML model on 1,420 questions, with 145 analyzed post-clustering. From this, 24 distinct challenges were identified and rated by 31 study participants. %
Additionally, five improvement suggestions were evaluated, and 12 more were uncovered via open-ended responses.

\new{Our findings suggest that VA language adoption is primarily blocked by lack of documentation, and usability and ecosystem limitations, highlighting key areas for improvement. For example, a challenge across all experience levels lies in designing and completing proof strategies, underscoring the need for better tool support and guidance. Additionally, the mismatch found between online discussions and reported challenges (e.g., frame specifications and proof termination clauses, were identified in Stack Overflow, but not confirmed as challenges), points to the importance of empirical user studies, as research based solely on online discussions may not fully capture the real experience of practitioners (or vice-versa).} %

This work can pave the way for future research on challenges and improvements, while also offering valuable insights to guide the design and/or enhancement of VA languages %
to support broader adoption. Future work includes expanding the dataset with further sources and carry out a large-size user study. In addition, newly identified challenges should be ranked based on relevance, in order to direct development towards support that can have a greater impact.

\vskip 0.1em
\noindent\noindent
{\bf Threats to Validity.} 
For \emph{topic modeling}, using question bodies provides more context but also introduces noise from off-topic discussions, error messages, and troubleshooting attempts, which may obscure the actual challenges. We mitigated this by reviewing both topic keywords and question bodies during classification. Additionally, Stack Overflow may lack sufficient coverage for some tools, limiting the generalizability of our results. Our mapping between languages and questions may also have missed relevant posts, reducing completeness.
In the \emph{survey}, the number of participants limited the ability to detect differences between experience levels. Also, the translation of challenges into affirmations may have lost important context, affecting participants’ understanding. We mitigated this by providing open-ended questions where participants could provide their opinions.

\section*{Acknowledgments}{
Alexandra Mendes was supported by an Amazon Research Award, Fall 2024. Carolina Carreira was financed by National Funds through the FCT - Fundação para a Ciência e a Tecnologia, I.P. (Portuguese Foundation for Science and Technology) within the project VeriFixer, with reference 2023.15557.PEX (DOI: 10.54499/2023.15557.PEX), and by funding from FCT under grant PRT/BD/153739/2021.
We thank João F. Ferreira for his valuable comments on this work.
}

\bibliographystyle{plain}  %
\bibliography{bib}         %

\end{document}